\documentstyle[preprint,aps]{revtex}
\begin{document}
\title{DYNAMICS OF THE SUBSETS OF NATURAL NUMBERS: A NURSERY RHYME OF CHAOS}
\author{ MOFAZZAL AZAM}
\address{THEORETICAL PHYSICS DIVISION, CENTRAL COMPLEX,
BHABHA ATOMIC RESEARCH CENTRE \\
TROMBAY, BOMBAY-400085 INDIA}
\maketitle
\vskip .8 in
\begin{abstract}
 We introduce a metric on the set of all subsets of natural
numbers which converts it into a cantor set.On this set endowed
with the metric, we introduce a very simple map which exhibits
chaotic behaviour.This simple map is almost like a nursery rhyme
of chaos.
\end{abstract}
\newpage
 The set of all subsets of natural numbers is a much larger set
than the set of natural numbers.In fact, it has the cardinality
of the set of real numbers.We shall denote the set of all subsets
of natural numbers by the symbol $\Sigma$, set of natural numbers
by {\bf N}, subsets of natural numbers by $\alpha$, $\beta$ etc.
\begin{eqnarray}
\alpha=(n_{1},n_{2},...,n_{k},...)
\nonumber\\
\beta=(m_{1},m_{2},...,m_{k},...)
\end{eqnarray}
Here $n_{k}$ and $m_{k}$ are natural numbers and $\alpha$, $\beta$
$\subset{\bf N}$ and $\in \Sigma$.On the set $\Sigma$ ,
we define symmetric set theoretic difference as
\begin{eqnarray}
\gamma \equiv (\alpha -\beta)
\nonumber\\
=(\alpha\bigcup \beta)\setminus (\alpha\bigcap \beta)
\nonumber\\
=(\alpha\setminus \beta)\bigcup (\beta \setminus \alpha)
\end{eqnarray}
We now define on the set $\Sigma$ a norm as follows
 (1) To the null set $\emptyset$, we associate zero i.e.,
$\| \emptyset \| =0$
 (2) To the set $ \alpha=(n_{1},n_{2},...,n_{k})$, we associate
 either
 \begin{eqnarray}
 \| \alpha \| = \sum_{n_{k} \in \alpha}(1/M^{n{_k}})
 \end{eqnarray}
 where $M \geq 2$.In fact, we take $M=2$.
  or
\begin{eqnarray}
\| S_{1}\| =\sum_{n_{k}\in \alpha}(1/n_{k}!)
\end{eqnarray}
In this paper, we will use the first definition of the norm though
we might as well have used the second one.
The distance between two subsets $\alpha$ and $\beta$ is given by
\begin{eqnarray}
\| (\alpha -\beta)\|  \equiv \| \gamma \|
\nonumber\\
=\| (\alpha \bigcup \beta)\setminus (\alpha \bigcap \beta)\|
\nonumber\\
=\| (\alpha \setminus \beta)\bigcup (\beta\setminus \alpha)\|
\end{eqnarray}
It is clear from the definitions that $\| \gamma \| =0$ imply
$\alpha=\beta$,  and the triangle inequality is easily satisfied.
On this set $\Sigma$ endowed with above mentioned metric, we introduce
the "Reduction by Unity" map $\Omega$, as follows
\begin{eqnarray}
(n_{1},n_{2},...,n_{k},..)\rightarrow (n_{1}-1,n_{2}-1,...,n_{k}-1,...)
\nonumber\\
\end{eqnarray}
if none of $n_{k}=1$.If one of $n_{k}$, say $n_{1}=1$, then
\begin{eqnarray}
(1,n_{2},...,n_{k},....)\rightarrow (n_{2}-1,...,n_{k}-1,....)
\end{eqnarray}
It is clear from the definition that the "Reduction by Unity" map
$\Omega$ is two-to-one.

 The above structures defined on the set  $\Sigma$ are very natural.
We explain this by the following physical example.Let us consider an
infinite number of boxes.Any number of boxes can be empty or filled
with particles.However, the filled boxes are such that no two boxes
have the same number of particles.The configuration of the particles
in the boxes can be represented by the sequences
$(n_{1},n_{2},...,n_{k},.....)$  where $n_{1},n_{2},...,n_{k},...$
are distinct positive integers.Note that we do not have zero as an
entry in the sequences even though there are any number of
empty boxes because for the process to be described in the following
these empty boxes are irrelevant except when all of them are empty.
We describe a process of evaporation of particles from the boxes
in the following way: at a time from each box a single particle is
evaporated.In other words, we define evaporation
per unit time
via the the map
$(n_{1},n_{2},...,n_{k},...)\rightarrow (n_{1}-1,n_{2}-1,...,n_{k}-1,...)$
(if $n_{1}=1$, then there is no entry at the location of $n_{1}-1$).
Note that this map is two-to-one.We arrive at the configuration
$(n_{1}-1,n_{2}-1,...,n_{k}-1,....)$ either from the evaporation of
the configuration $(n_{1},n_{2},...,n_{k},....)$  or
$(1,n_{1},n_{2},...,n_{k},....)$.The dynamics of the evaporation can
not be analysed any more unless we introduce some kind of distance or
metric on the set of all possible configuration of the particles in the
boxes.This will tell us how far or close particle configurations are
from each other.An adhoc mathematical definition would do the job.
However, we shall try to be as close as possible to a physical situation.
It is natural to assume that two configurations with similar number of
particles in the boxes are closer to each other than those with
different number of particles.The more the dissimilar number of
entries in the two configurations the further they are.Another assumption
we would like to make is that two configurations differing from each other
in very large entries are not too far from each other.This sounds a bit
unatural.However, it is required mathematically.Note that the
configuration $\emptyset$, corresponding to "no particle" in any of
the boxes also belongs to the set of all possble configurations.
Therefore, the distance of any configuration from the null configuration
$\emptyset$ essentially defines a norm.Now if the distance between two
configurations keep increasing with larger entries of particles in
the boxes there will be problem with convergence of the norm.Therefore,
the two requirement mentioned above are necessary for defining a
proper metric.This metric introduces an interesting scenario for the
evaporation process: there is a temporal scale separation- the short time
and long time behaviour of the evaporation is different
in the sense that two generic configurations which
stay close to each other in short time separate exponentially
from each other in the long time.In fact, what will
be proven in this paper is that the set of
possible configurations of particles
in the boxes (i.e., the set of all subsets of natural numbers $\Sigma$)
is a Cantor set and the evaporation process (i.e.,the "Reduction by Unity"
map $\Omega$ on $\Sigma$) is chaotic.

 A set is called a Cantor set if it is closed,
totally disconnected  and
is a perfect set."Closed" means every sequence of of its'
points converges
to some point within the set.
"Totally diconnected" means it contains no interval.A perfect set
means every point in it is an accumulation point.The classical example
of a Cantor set is the "Cantor middle-thirds" set.The "Cantor
middle-thirds" set is nicely described by Cantor map
\begin{eqnarray}
F(x)=3x    \quad \mbox{for }\quad  -\infty < x \leq 1/2
\nonumber\\
    =3(1-x) \quad \mbox{for }\quad  1/2 < x \leq \infty
\end{eqnarray}
on the real line {\bf R}.Note that most points of the real
line {\bf R} go to $-\infty$ under the iteration of the Cantor
map.However, there exists a set which is a subset of the
interval {\bf I=[0,1]} that does stay within the interval {\bf I}
and is invariant under the iteration of the map F(x).This is
precisely the "Cantor middle-thirds" set.The Cantor map F(x) on
this set is chaotic \cite{1,2}.This means that,

(1) The map has the property of sensitive dependence on initial
 conditions.

(2) It has dense set of periodic orbits.

(3) It has a topologically transitive (chaotic) trajectory i.e.,
a trajectory which goes arbitrary close to every point in the set.

 We shall proof that the set of all subsets
of natural numbers, $\Sigma$ endowed with the metric introduced
in begining of paper, is {\bf homeomorphic} to the Cantor set.In
other words there exists,
between the "Cantor middle-thirds" set and the set of all subsets
of natural numbers, $\Sigma$, a mapping which is
one-one , onto and continuous along with the
the inverse.We shall also proof that
the Cantor map on Cantor set is topologically conjugate to the
"Reduction by Unity" map on $\Sigma$.

From the definition of the Cantor map, note that $F'(x)=3$, and
this means that the map is expansive.Moreover, $F^{-1}(I)$, where
$I=[0,1]$,consists of two subintervals  $I_{0}=[0,1/3]$
and $I_{1}=[2/3 ,1]$.Also for any subinerval $J\subset I$,
$F^{-1}(J)$ consists of two subintervals, one in $I_{0}$ and
the other in $I_{1}$.The length of the subinterval $J$ is
3-times the length of either of the subintervals $F^{-1}(J)$.
Also, $F^{-n}(I_{0})$ and $F^{-n}(I_{1}) \rightarrow 0$ as
$3^{-n}$.
These properties of the Cantor map can be used to establish a
homeomorphism between $\Lambda$, the Cantor set and
the set $\Sigma$.

Let us
construct the itinerary S(x) of a point x under the Cantor
map F(x),
 \begin{eqnarray}
 S(x)=(n_{1},n_{2},...,n_{k},....)
 \end{eqnarray}
The symbols $n_{j}$ are introduced in the following way:
$n_{j}\in S(x)$ if $F^{n_{j}-1}(x)\in I_{0}$.
It will be argued
that S(x) defined above is a homeomorphism.Firstly, it is
obvious that to every point $x\in \Lambda $, where $\Lambda$ is
the Cantor set, there is sequenc S(x) i.e., a subset of $\Sigma$.

Now we proof that to every subset of $\Sigma$ there is a point
$x\in \Lambda$.
Let $\eta=(1,2,...,n)$ and $ \xi ,\zeta \subset \eta$.
$\xi \bigcup \zeta=\eta$ and $ \xi \bigcap \zeta =\emptyset$,
the nullset.
Define  $\xi_{1}$ and $\zeta_{1}$ as two sets obtained
from $\xi$ and $\zeta$ by "Unit Reduction map".Therefore,
$\xi_{1} \bigcup \zeta_{1}=\eta_{1}$, where
$\eta_{1}=(1,2,...,n-1)$ and
$\xi_{1} \bigcap \zeta_{1}=\emptyset $, the nullset.Define,
\begin{eqnarray}
I^{n}_{\xi,\zeta}=\{ x\in I\quad\mid F^{k-1}(x)\in I_{0}\quad \mbox{for}\quad
\nonumber\\
k\in \xi, F^{k-1}(x)\in I_{1}\quad \mbox{for}\quad k\in \zeta \}
\nonumber\\
=(\bigcap_{k\in \xi} F^{-k+1}(I_{0}))(\bigcap_{k\in \zeta} F^{-k+1}(I_{1}))
\end{eqnarray}

Let us proof that $I^{n}_{\xi, \zeta}$ is a connected interval.
First note that  integer, $1\in \xi$ or $1\in \zeta$, therefore,
\begin{eqnarray}
I^{n}_{\xi,\zeta}
=(I_{0}\quad \mbox{or}\quad I_{1})\bigcap F^{-1}(I^{n-1}_{\xi_{1}, \zeta_{1}})
\end{eqnarray}
By induction, we assume that $I^{n-1}_{\xi_{1}, \zeta_{1}}$
is non-empty connected subinterval.Therefore,
$F^{-1}(I^{n-1}_{\xi_{1}, \zeta_{1}}) $ consists of two subintervals
one in $I_{0}$ and the other in $I_{1}$.Therefore,
$(I_{0}\quad \mbox{or}\quad I_{1})\bigcap F^{-1}(I^{n-1}_{\xi_{1}, \zeta_{1}})$
consists of a single interval.These intervals are nested.To prove this, we
introduce slightly different notations.
\begin{eqnarray}
I^{n}=\{ x \in I\quad\mid F^{k-1}(x)\in I_{0}\quad \mbox{for}\quad
\nonumber\\
k\in \xi, F^{k-1}(x)\in I_{1}\quad \mbox{for}\quad k\in \zeta \}
\nonumber\\
I^{n-1}=[x\in I\quad\mid F^{k-1}(x)\in I_{0}\quad \mbox{for}\quad
\nonumber\\
k\in \xi _{2}, F^{k-1}(x)\in I_{1} \mbox{for} k\in \zeta_{2}
\end{eqnarray}
$\xi_{2}=\xi$ if the integer n is not an entry in the set $\xi$,
otherwise $\xi_{2}$ is obtained from the set $\xi$ only by dropping
the integer n.$\zeta_{2}$ is defined similarly.
Now,
\begin{eqnarray}
\mbox{if}\quad n\in \xi\quad \mbox{then}
I^{n}=I^{n-1}\bigcap \{x\in I\quad\mid F^{n-1}(x)\in I_{0 } \}
\nonumber\\
\mbox{if}\quad n\in \zeta\quad \mbox{then}
I^{n}=I^{n-1}\bigcap \{ x\in I\quad\mid F^{n-1}(x)\in I_{1} \}
\end{eqnarray}
Therefore, $I^{n}\subset I^{n-1}$
This implies that
$\bigcap_{n} I^{n}$
is a nested sequences of subintervals and, therefore, it is non-empty.
The subintervals  $ I^{n}$ are contained in $I_{0}$ or $I_{1}$ and
their lengths tend to zero as $n\rightarrow \infty$, and, therefore,
the intersection is a unique point.
Now, when $n\rightarrow \infty$, $I^{n}_{\xi,\zeta}$ is totally
given by $\xi$, and $\zeta$ is just its' complement in the set of natural
numbers.Therefore, to every subset $\xi \in \Sigma$, the set of all subsets
natural numbers, there exists a unique point
$ x\in \Lambda$, the Cantor set.This implies that the itinerary,
S(x) is onto.

\par That distinct points of Cantor set, $\Lambda$ correspond
to different
subsets of natural numbers can be seen from the fact that the
subintervals
$I^{n}(x)$,containing the point $x$, ( constructed in the previous
section) tend to zero as
$3^{-n+1}$ when $n\rightarrow \infty$.For $x\neq y$, consider
neighbourhoods $\aleph_{x}$ and $\aleph_{y}$ of $x$ and $y$ such
that $\aleph_{x} \bigcap \aleph_{y}=\emptyset$.Since the length
of the subintervals $I^{n} \rightarrow 0$ as $3^{-n+1}$,
there exist an $n_{\star}$, such that for $n > n_{\star}$
$I^{n}(x) \in \aleph_{x}$ and $I^{n}(y) \in \aleph_{y}$.

\par In order to proof continuity, let us choose an $\epsilon$.Now
choose $n$ such that $1/2^{n}< \epsilon$.Let $x\in \Lambda$
and $S(x)=(n_{1},n_{2},...,n_{k},....)$.Consider all subintervals
of the form $I^{n}_{\xi, \zeta}$ with $n$ fixed but $\xi$ and $\zeta$
varying.
There are $2^n + 1$ such intervals.
These subintervals are disjoint and $\Lambda$ is contained in
the union of them.Hence, we can choose a $\delta$ such that
$\mid x-y\mid\quad < \delta$, and $y\in \Lambda$, imply that $y\in I^{n}(x)$.
Therefore, itinerary of the of the points ,
$S(x)$ and $S(y)$ agree on $n$ entries.Now, from the metric
on $\Sigma$ it follows that
\begin{eqnarray}
\| (S(x)-S(y))\| \leq 1/2^{n} < \epsilon
\end{eqnarray}
This proves continuity.Similarly, one can prove continuity of
the inverse.Therefore, the itinerary
$S(x)$ of points $x\in \Lambda$ is a homeomorphism between the
Cantor set, $\Lambda$ and the set of all subsets of natural number
$\Sigma$.A set homeomorphic to a Cantor set is itself a Cantor set.
This proves the main claim of the paper.

\par Now, we describe some
interesting and important corrolaries of this result.
Now, we prove that the "Reduction by Unity" map $\Omega$
on set $\Sigma$ and Cantor map F(x) on set $\Lambda$ are topologically
conjugate to each other.In other words,
\begin{eqnarray}
S\circ F(x)=\Omega\circ S(x)
\end{eqnarray}
\par A point $x \in \Lambda$ is uniquely determined by intersection
of nested sequence of intervals $\bigcap_{n} I^{n}$
These subintervals as argued earlier in the paper are determined
by the itinerary $S(x)$.
\begin{eqnarray}
I^{n}_{\xi,\zeta}
\nonumber\\
=(I_{0} \quad \mbox{or} \quad I_{1}) \bigcap
F^{-1}(I^{n-1}_{\xi_{1}, \zeta_{1}})
\end{eqnarray}
As explained earlier, $\xi_{1}, \zeta_{1}$ are obtained
from $\xi$ and $\zeta$ by unit reduction map.
Therefore,
\begin{eqnarray}
F(x) = F(I^{n}_{\xi,\zeta},\quad \mbox{when}\quad n\rightarrow \infty)
\nonumber\\
=I^{n-1}_{\xi_{1}, \zeta_{1}}\quad \mbox{when}\quad n\rightarrow \infty
\end{eqnarray}
From here it follows that
\begin{eqnarray}
S\circ F(I^{n}_{\xi,\zeta},\quad \mbox{when}\quad n\rightarrow \infty )
\nonumber\\
=S(I^{n-1}_{\xi_{1}, \zeta_{1}},\quad \mbox{when}\quad n\rightarrow \infty )
\nonumber\\
=(n_{1}-1,n_{2}-1,...,n_{k}-1,....)
\nonumber\\
=\Omega (n_{1},n_{2},...,n_{k},....)
\nonumber\\
=\Omega \circ S(x)
\end{eqnarray}
The "Reduction by Unity"  map $\Omega$ on
the set of all subsets of natural numbers, $\Sigma$ is topologically
conjugate to the Cantor map $F(x)$ on the Cantor set, $\Lambda$.This
also proves that the evaporation process introduced earlier is also
chaotic.This hypothetical evaporation process is a very generic
example of chaos and, therefore, before we end the paper we mention
the following interesting corrollary.Let us construct a
set of functions in the following way: to every configuration
$\alpha=(n_{1},n_{2},...,n_{k},....)$, we associate
a function $h_{\alpha}(x)$
\begin{eqnarray}
h_{\alpha}(x) &=& \exp(x)-\sum_{k\in \alpha}\frac{x^{k-1}}{(k-1)!}
\nonumber\\
&=&\sum_{k\in \beta}\frac{x^{k-1}}{(k-1)!}
\end{eqnarray}
where $\alpha \bigcup \beta = {\bf N}$ and $\alpha \bigcap \beta = \emptyset$ .
When $\beta = \emptyset$,
$\alpha ={\bf N}$, and therefore, according to the defination
above $h_{\emptyset}(x)=0$ and corresponds to set $\emptyset\in \Sigma$.
It is clear from the defination that functions $h(x)$ are obtained
from the exponential series
\begin{eqnarray}
\exp(x)=1+x+\frac{x^{2}}{2!}+....+\frac{x^{k}}{k!}+....
\nonumber\\
\end{eqnarray}
by randomly dropping finite or infinite number of terms.We deote
this set functions as $\Xi$.We introduce on the set of functions,
$\Xi$ a metric in the following way,
\begin{eqnarray}
\|(h_{\alpha}(x)-h_{\beta}(x))\|
=\sum_{n=0}^{\infty} \frac{\mid h_{\alpha}^{(n)}(x)-h_{\beta}^{(n)}(x)\mid}
{2^{n}} \Big|_{x=0}
\end{eqnarray}
here $\alpha$ and $\beta$ are any subset in $\Sigma$ and
$h_{\alpha}^{(n)}(x)$
is the $n^{th}$ derivative of $h_{\alpha}(x)$.It is clear that
$\| (h_{\alpha}(x)-h_{\beta}(x)) \| = 0$ imply
$h_{\alpha}(x)=h_{\beta}(x)$ and the triangle inequality is
satisfied.It is very easily proven that the set of randomly diluted
exponetial functions, $\Xi$ is a Cantor set in the above metric and
the "Derivative" map on the functions of $\Xi$ is chaotic.
\par I thank Agha A. Ali of Theoretical Physics Group, Tata Institute
of Fundamental Research, Bombay for discussion and many valuable
suggestions.


\begin{thebibliography}{99}
\bibitem{1}  R.L. Devaney-Introduction to Chaotic
Dynamics, 1986, The  Benjamin Cummings Publishing Co. Inc.,
Menlo Park, California, Reading, Massachusetts.

\bibitem{2} J. Guckenheimer and P. Holmes-Nonlinear
Oscillations, Dynamical Systems, and Bifurcations of Vector
Fields, 1983, Springer-Verlag,New York-Berlin-Heidelberg-Tokyo.
\end{thebibliography}
\end{document}